\documentclass{article}
\usepackage{spconf,amsmath,graphicx}

\ninept
\title{Joint Demosaicing and Denoising with Double Deep Image Priors}
\name{Taihui Li ${^1}$\sthanks{Work performed while at Sony Corporation of America.} \quad Anish Lahiri ${^2}$ \quad Yutong Dai ${^2}$ \quad Owen Mayer ${^2}$}

\address{${^1}$ Computer Science and Engineering, University of Minnesota, Minneapolis, USA\\
${^2}$ Sony Corporation of America, R\&D US Laboratory, San Jose, USA}

\usepackage{graphicx,amsmath,amsfonts,amsthm,amscd,amssymb,bm,url,color,latexsym}
\usepackage{physics}
\allowdisplaybreaks
\usepackage[capitalize,nameinlink]{cleveref}

\renewcommand{\mathbf}{\boldsymbol}

\newcommand{\mb}{\mathbf}

\newcommand{\bb}{\mathbb}

\newcommand{\R}{\bb R}

\newcommand{\wh}{\widehat}

\usepackage{booktabs}
\usepackage{multirow}
\usepackage{makecell}

\usepackage[numbers]{natbib}

\begin{document}
\newcommand{\red}[1]{ \textcolor{red}{#1}}
%
\maketitle
\begin{abstract}
Demosaicing and denoising of RAW images are crucial steps in the processing pipeline of modern digital cameras. As only a third of the color information required to produce a digital image is captured by the camera sensor, the process of demosaicing is inherently ill-posed. The presence of noise further exacerbates this problem. Performing these two steps sequentially may distort the content of the captured RAW images and accumulate errors from one step to another. Recent deep neural-network-based approaches have shown the effectiveness of joint demosaicing and denoising to mitigate such challenges. However, these methods typically require a large number of training samples and do not generalize well to different types and intensities of noise. In this paper, we propose a novel joint demosaicing and denoising method, dubbed JDD-DoubleDIP, which operates directly on a single RAW image without requiring any training data. We validate the effectiveness of our method on two popular datasets---Kodak and McMaster---with various noises and noise intensities. The experimental results show that our method consistently outperforms other compared methods in terms of PSNR, SSIM, and qualitative visual perception. 
\end{abstract}
%
%


\section{Introduction}\label{sec:introduction}




A RAW image (a.k.a. mosaic image) is sensor data directly captured by digital cameras. In RAW images, only a third of the color information required to produce a high-quality full-color RGB image is available. Hence, demosaicing is necessary to interpolate the missing color components. However, this is inherently an ill-posed problem~\cite{DBLP_journals_tog_Durand16a,DBLP_conf_cvpr_LiuJL020} and the presence of noise in RAW images due to various factors (e.g., Poisson noise due to lighting conditions, Gaussian noise from electronics, etc.)  further exacerbates this problem~\cite{DBLP_reference_vision_Hasinoff14}. Traditionally, demosaicing and denoising are handled sequentially in the camera processing pipeline, but this may lead to content distortion in images and the accumulation of errors from one processing step to another. Furthermore, determining the optimal processing order also becomes a challenge~\cite{DBLP_conf_iccp_QianWGDHGR22,DBLP_conf_cvpr_JinFM20}.

\begin{figure}[!tb]
\centering
\includegraphics[width=0.48\textwidth]{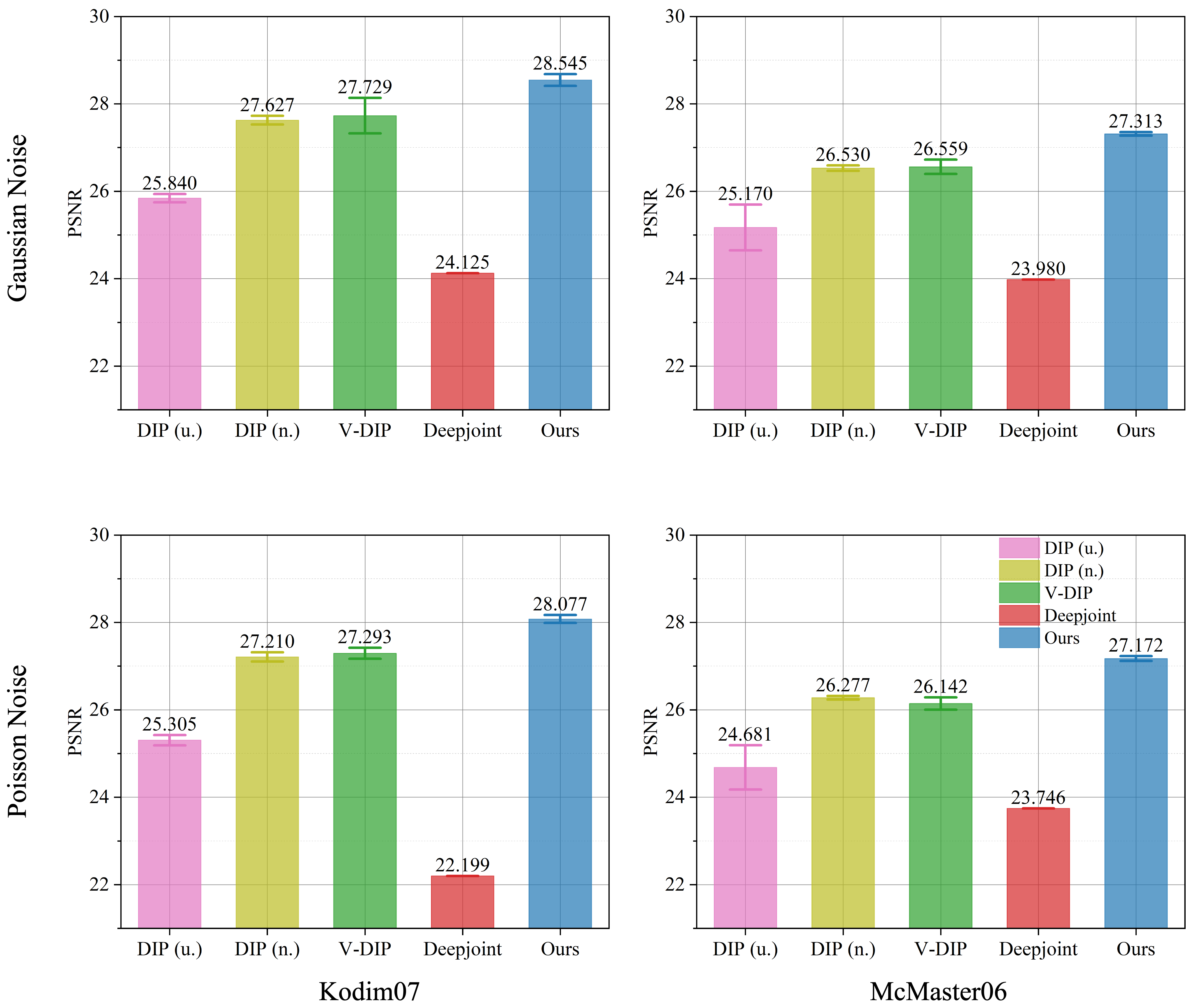}
\vspace{-20pt}
\caption{The PSNR comparisons of demosaicing and denoising on RAW images---Kodim07 (first column) and McMaster06 (second column)---with Gaussian noise $\sigma=30$ (first row) and Poisson noise $\lambda=25$ (second row). We provide comprehensive comparisons in~\cref{tab_gaussin_noise} and~\cref{tab_poisson_noise}.}
\vspace{-20pt}
\label{fig_fig1}
\end{figure}


In contrast, adopting a joint demosaicing and denoising (JDD) strategy can naturally overcome the aforementioned challenges~\cite{DBLP_journals_tip_HirakawaP06,DBLP_journals_tog_Durand16a}. Classical methods for JDD include the specialized design of filters with constraints~\cite{DBLP_journals_tip_HirakawaP06} and the use of certain heuristics such as total variation minimization~\cite{DBLP_conf_icip_CondatM12}, self-similarity~\cite{heide_flexisp_2014}, learned
nonparametric random fields~\cite{DBLP_journals_tip_KhashabiNJF14}, and sequential energy minimization~\cite{DBLP_conf_iccp_KlatzerHKP16}. Recently, with the resurgence of deep neural networks (DNNs), several studies~\cite{DBLP_journals_tog_Durand16a,DBLP_conf_iccv_EhretDAF19,DBLP_conf_cvpr_LiuJL020,DBLP_journals_sensors_ParkLJY20} have shown the benefits of DNN-based JDD over traditional methods. One main stream among these methods is data-driven JDD. Gharbi et al.~\cite{DBLP_journals_tog_Durand16a} cast JDD as a supervised learning problem and use a convolutional neural network to directly learn a mapping between noisy RAW images and full-color RGB images. Ehret et al.~\cite{DBLP_conf_iccv_EhretDAF19} propose a mosaic-to-mosaic training strategy that learns demosaicing and denoising on RAW images only, without full-color ground-truth RGB images. Liu et al.~\cite{DBLP_conf_cvpr_LiuJL020} design an additional branch to estimate the green channel and then use the estimated green channel as a guide to recover all missing values. Though these methods have significantly improved the demosaicing and denoising performance, they require massive training data to learn the models well. It is not only expensive to acquire these data sets, but there are no ground-truth data in practice. Worse, once a DNN model has been trained on a particular noise and noise intensity, it does not generalize well to other noise and noise intensities.

\begin{figure*}[!htpb]
\centering
\includegraphics[width=0.6\textwidth]{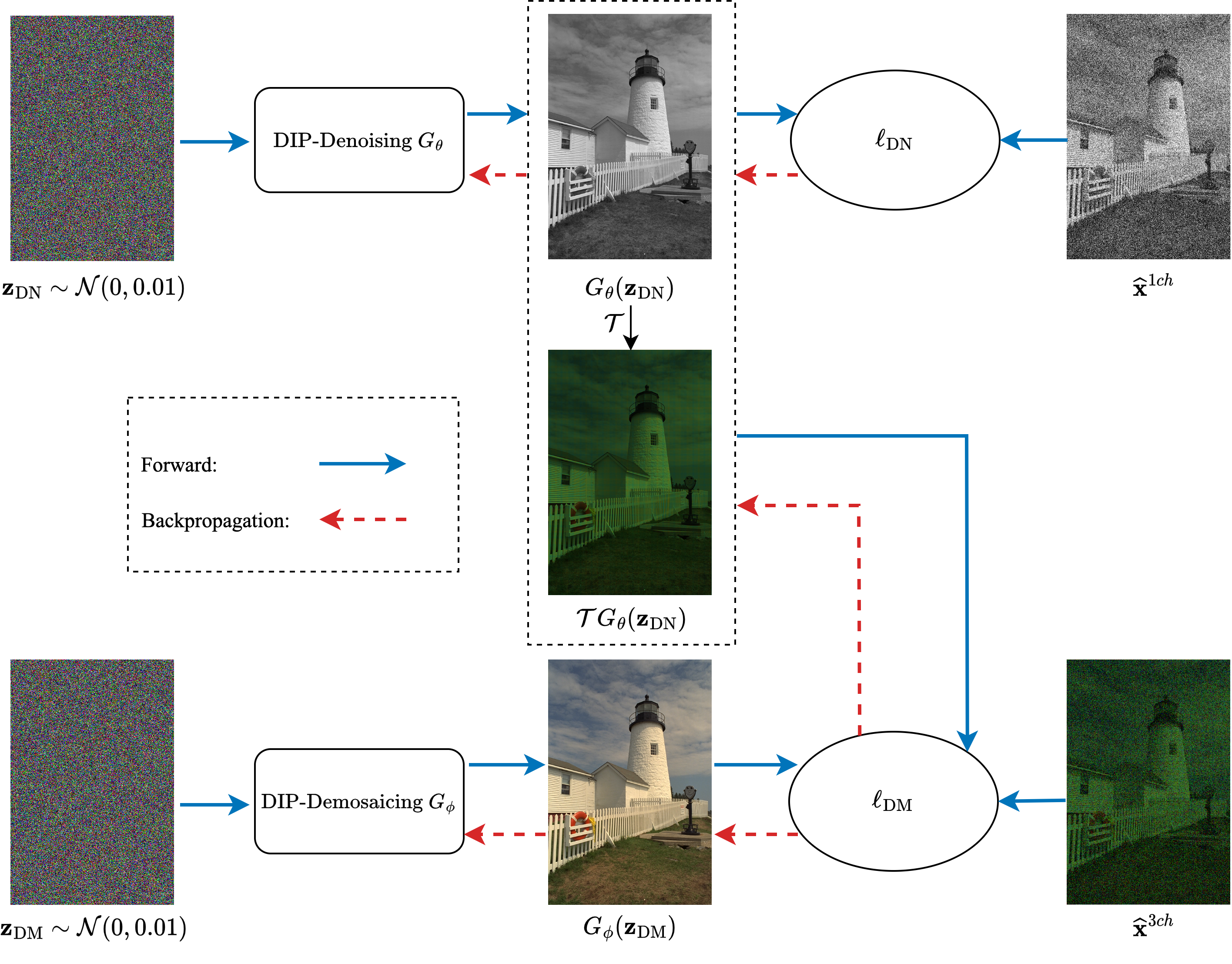}
\vspace{-15pt}
\caption{Framework of JDD-DoubleDIP, consisting of a~\emph{DIP-Denoising} $G_{\mb \theta}$ and a~\emph{DIP-Demosaicing} $G_{\mb \phi}$. $G_{\mb \theta}$ is used to denoise the given noisy RAW image and provide ``clean" guidance to $G_{\mb \phi}$; while $G_{\mb \phi}$ attempts to recover the desired high-quality full-color RGB image based on 1)~the given noisy RAW image $\wh{\mb x}^{3ch}$ and 2) the denoised output of $G_{\mb \theta}$. We join these two branches by combining their respective losses $\ell_{\text{JDD-DoubleDIP}} = \sqrt{\ell_{\text{DN}}} + \sqrt{\ell_{\text{DM}}}$ and train them simultaneously. By doing so, these two branches collaborate with each other, yielding better demosaicing and denoising results.}\vspace{-15pt}
\label{fig_framework}
\end{figure*}

This motivates the need for developing single-instance JDD methods. 
The recent deep image prior (DIP) method~\cite{DBLP_conf_cvpr_UlyanovVL18} demonstrates that the neural network itself can serve as an implicit prior for natural images to solve image restoration problems by fitting a single noisy observation. Although DIP has garnered growing interest recently, its application in RAW image demosaicing remains relatively scarce. Park et al.~\cite{DBLP_journals_sensors_ParkLJY20} for the first time present a DIP-based approach (V-DIP) for RAW image demosaicing and denoising, obviating the need for training data. However, V-DIP has an additional optimization objective to update targets and formulates the problem as image inpainting without explicitly considering denoising.


To address the limitations of V-DIP, we propose a dual-branch model, dubbed~\emph{JDD-DoubleDIP}, which utilizes information from a denoising DIP branch to guide the joint demosaicing and denoising objective. To do this, we first adopt a DIP dedicated to account for noise removal (\emph{denoising}). Then, we employ a second DIP which jointly performs denoising and missing-value interpolation (\emph{demosaicing}), and is guided by the output from the denoising DIP. We join these two DIPs by amalgamating their respective loss functions into a single optimization objective and update their parameters simultaneously. By doing so, we have not only explicitly induced denoising guidance information into the model, but we also enabled these two DIPs to collaborate with each other for better performance.~\cref{fig_fig1} shows the superiority of our model over other models. 

\begin{table*}[!htpb]
\caption{The quantitative results of demosaicing and denoising of RAW images with Gaussian noise. \textbf{Deepjoint}: we use the maximum noise intensity of the pretrained model as input. \textbf{$\text{Deepjoint}^{*}$}: we use the exact noise intensity of the noisy RAW image as input. \textbf{$\text{Ours}^{+}$}: we smooth the output of our model by using running average: $\mb x^{3ch}_{\text{smooth}} = 0.99*\mb x^{3ch}_{\text{smooth}} + 0.01*G_{\phi}(\mb z_{\text{DM}}).$} 
\vspace{-8pt}
\label{tab_gaussin_noise}
\begin{center}
\setlength{\tabcolsep}{1.15mm}{

\begin{tabular}{c l l l l l l| l l l l l}

&
& 
\multicolumn{5}{c}{\scriptsize{PSNR $\uparrow$}} &
\multicolumn{5}{c}{\scriptsize{SSIM $\uparrow$}}
\\
\hline

\hline

\hline
\scriptsize{Dataset}

& \scriptsize{Method}

& \scriptsize{$\sigma$=10}
& \scriptsize{$\sigma$=20}
& \scriptsize{$\sigma$=30}
& \scriptsize{$\sigma$=50}
& \scriptsize{$\sigma$=70}

& \scriptsize{$\sigma$=10}
& \scriptsize{$\sigma$=20}
& \scriptsize{$\sigma$=30}
& \scriptsize{$\sigma$=50}
& \scriptsize{$\sigma$=70}
\\
\hline

\hline
\multirow{7}{*}{\rotatebox{90}{\scriptsize{Kodak}}}

&\scriptsize{DIP (u.)}
&\scriptsize{29.383} \tiny{(0.038)}
&\scriptsize{26.521} \tiny{(0.021)}
&\scriptsize{24.786} \tiny{(0.017)}
&\scriptsize{22.707} \tiny{(0.021)}
&\scriptsize{21.369} \tiny{(0.014)}

&\scriptsize{0.811} \tiny{(0.0012)}
&\scriptsize{0.682} \tiny{(0.0010)}
&\scriptsize{0.595} \tiny{(0.0021)}
&\scriptsize{0.494} \tiny{(0.0028)}
&\scriptsize{0.443} \tiny{(0.0033)}
\\

&\scriptsize{DIP (n.)}
&\scriptsize{30.061} \tiny{(0.028)}
&\scriptsize{27.594} \tiny{(0.071)}
&\scriptsize{26.104} \tiny{(0.015)}
&\scriptsize{24.132} \tiny{(0.009)}
&\scriptsize{22.594} \tiny{(0.014)}

&\scriptsize{0.840} \tiny{(0.0009)}
&\scriptsize{0.748} \tiny{(0.0030)}
&\scriptsize{0.684} \tiny{(0.0010)}
&\scriptsize{0.589} \tiny{(0.0015)}
&\scriptsize{0.516} \tiny{(0.0009)}
\\

&\scriptsize{V-DIP}
&\scriptsize{29.664} \tiny{(0.154)}
&\scriptsize{27.393} \tiny{(0.037)}
&\scriptsize{25.872} \tiny{(0.022)}
&\scriptsize{23.792} \tiny{(0.027)}
&\scriptsize{22.269} \tiny{(0.043)}

&\scriptsize{0.834} \tiny{(0.0053)}
&\scriptsize{0.759} \tiny{(0.0019)}
&\scriptsize{0.703} \tiny{(0.0005)}
&\scriptsize{0.614} \tiny{(0.0014)}
&\scriptsize{0.544} \tiny{(0.0013)}
\\





&\scriptsize{Deepjoint}
&\scriptsize{30.040} \tiny{(N/A)}
&\scriptsize{29.998} \tiny{(N/A)}
&\scriptsize{23.796} \tiny{(N/A)}
&\scriptsize{17.494} \tiny{(N/A)}
&\scriptsize{14.558} \tiny{(N/A)}

&\scriptsize{0.792} \tiny{(N/A)}
&\scriptsize{0.818} \tiny{(N/A)}
&\scriptsize{0.482} \tiny{(N/A)}
&\scriptsize{0.231} \tiny{(N/A)}
&\scriptsize{0.144} \tiny{(N/A)}
\\

&\scriptsize{$\text{Deepjoint}^{*}$}
&\scriptsize{33.222} \tiny{(N/A)}
&\scriptsize{29.995} \tiny{(N/A)}
&\scriptsize{27.828} \tiny{(N/A)}
&\scriptsize{24.068} \tiny{(N/A)}
&\scriptsize{19.665} \tiny{(N/A)}

&\scriptsize{0.902} \tiny{(N/A)}
&\scriptsize{0.818} \tiny{(N/A)}
&\scriptsize{0.717} \tiny{(N/A)}
&\scriptsize{0.447} \tiny{(N/A)}
&\scriptsize{0.165} \tiny{(N/A)}
\\

&\scriptsize{Ours}
&\scriptsize{30.758} \tiny{(0.046)}
&\scriptsize{28.493} \tiny{(0.055)}
&\scriptsize{26.987} \tiny{(0.034)}
&\scriptsize{24.776} \tiny{(0.016)}
&\scriptsize{23.043} \tiny{(0.026)}

&\scriptsize{0.865} \tiny{(0.0014)}
&\scriptsize{0.788} \tiny{(0.0009)}
&\scriptsize{0.730} \tiny{(0.0008)}
&\scriptsize{0.648} \tiny{(0.0018)}
&\scriptsize{0.593} \tiny{(0.0014)}
\\

&\scriptsize{$\text{Ours}^{+}$}
&\scriptsize{31.149} \tiny{(0.048)}
&\scriptsize{28.919} \tiny{(0.062)}
&\scriptsize{27.382} \tiny{(0.037)}
&\scriptsize{25.014} \tiny{(0.019)}
&\scriptsize{23.119} \tiny{(0.030)}

&\scriptsize{0.874} \tiny{(0.0018)}
&\scriptsize{0.803} \tiny{(0.0010)}
&\scriptsize{0.746} \tiny{(0.0002)}
&\scriptsize{0.662} \tiny{(0.0020)}
&\scriptsize{0.603} \tiny{(0.0020)}
\\
\hline

\hline
\multirow{7}{*}{\rotatebox{90}{\scriptsize{McMaster }}}

&\scriptsize{DIP (u.)}
&\scriptsize{29.921} \tiny{(0.029)}
&\scriptsize{26.703} \tiny{(0.031)}
&\scriptsize{24.760} \tiny{(0.036)}
&\scriptsize{22.384} \tiny{(0.060)}
&\scriptsize{20.354} \tiny{(0.013)}

&\scriptsize{0.828} \tiny{(0.0008)}
&\scriptsize{0.701} \tiny{(0.0013)}
&\scriptsize{0.610} \tiny{(0.0022)}
&\scriptsize{0.510} \tiny{(0.0028)}
&\scriptsize{0.418} \tiny{(0.0039)}
\\

&\scriptsize{DIP (n.)}

&\scriptsize{30.843} \tiny{(0.016)}
&\scriptsize{28.065} \tiny{(0.026)}
&\scriptsize{26.155} \tiny{(0.005)}
&\scriptsize{23.450} \tiny{(0.008)}
&\scriptsize{21.391} \tiny{(0.013)}

&\scriptsize{0.859} \tiny{(0.0001)}
&\scriptsize{0.768} \tiny{(0.0008)}
&\scriptsize{0.690} \tiny{(0.0007)}
&\scriptsize{0.568} \tiny{(0.0009)}
&\scriptsize{0.481} \tiny{(0.0018)}
\\

&\scriptsize{V-DIP}
&\scriptsize{30.887} \tiny{(0.036)}
&\scriptsize{27.902} \tiny{(0.036)}
&\scriptsize{25.844} \tiny{(0.039)}
&\scriptsize{23.146} \tiny{(0.056)}
&\scriptsize{21.192} \tiny{(0.041)}

&\scriptsize{0.863} \tiny{(0.0007)}
&\scriptsize{0.777} \tiny{(0.0007)}
&\scriptsize{0.703} \tiny{(0.0011)}
&\scriptsize{0.590} \tiny{(0.0022)}
&\scriptsize{0.508} \tiny{(0.0019)}
\\





&\scriptsize{Deepjoint}
&\scriptsize{30.497} \tiny{(N/A)}
&\scriptsize{30.146} \tiny{(N/A)}
&\scriptsize{24.083} \tiny{(N/A)}
&\scriptsize{17.987} \tiny{(N/A)}
&\scriptsize{14.890} \tiny{(N/A)}

&\scriptsize{0.835} \tiny{(N/A)}
&\scriptsize{0.823} \tiny{(N/A)}
&\scriptsize{0.510} \tiny{(N/A)}
&\scriptsize{0.257} \tiny{(N/A)}
&\scriptsize{0.164} \tiny{(N/A)}
\\

&\scriptsize{$\text{Deepjoint}^{*}$}
&\scriptsize{33.142} \tiny{(N/A)}
&\scriptsize{30.140} \tiny{(N/A)}
&\scriptsize{27.953} \tiny{(N/A)}
&\scriptsize{24.012} \tiny{(N/A)}
&\scriptsize{19.495} \tiny{(N/A)}

&\scriptsize{0.903} \tiny{(N/A)}
&\scriptsize{0.834} \tiny{(N/A)}
&\scriptsize{0.744} \tiny{(N/A)}
&\scriptsize{0.486} \tiny{(N/A)}
&\scriptsize{0.207} \tiny{(N/A)}
\\

&\scriptsize{Ours}
&\scriptsize{31.544} \tiny{(0.060)}
&\scriptsize{28.739} \tiny{(0.028)}
&\scriptsize{26.880} \tiny{(0.013)}
&\scriptsize{24.075} \tiny{(0.038)}
&\scriptsize{21.906} \tiny{(0.020)}

&\scriptsize{0.876} \tiny{(0.0009)}
&\scriptsize{0.793} \tiny{(0.0011)}
&\scriptsize{0.726} \tiny{(0.0013)}
&\scriptsize{0.623} \tiny{(0.0020)}
&\scriptsize{0.552} \tiny{(0.0012)}
\\

&\scriptsize{$\text{Ours}^{+}$}
&\scriptsize{32.043} \tiny{(0.056)}
&\scriptsize{29.198} \tiny{(0.024)}
&\scriptsize{27.228} \tiny{(0.015)}
&\scriptsize{24.225} \tiny{(0.028)}
&\scriptsize{21.906} \tiny{(0.022)}

&\scriptsize{0.884} \tiny{(0.0007)}
&\scriptsize{0.804} \tiny{(0.0008)}
&\scriptsize{0.738} \tiny{(0.0012)}
&\scriptsize{0.634} \tiny{(0.0017)}
&\scriptsize{0.561} \tiny{(0.0018)}
\\
\hline

\hline

\hline
\end{tabular}
}
\end{center}\vspace{-20pt}
\end{table*}

\section{Our Method}\label{sec:method}

\subsection{Background: Deep Image Prior}
Given a single noisy observation $\mb y$, deep image prior (DIP)~\cite{DBLP_conf_cvpr_UlyanovVL18} uses a structured deep neural network $G_{\mb \theta}(\cdot)$ parameterized by $\mb \theta$ to fit this noisy observation and restore its corresponding clean image $\mb x$. Specifically, DIP solves the following optimization problem:
\begin{align} \label{eq:data_fitting_dip}
    \min_{\mb \theta} \; \ell(\mb y, f \circ G_{\mb \theta}(\mb z))
\end{align}
where $\circ$ denotes function composition, $f$ represents the forward measurement operator (e.g., $f$ is an identity operator for image denoising), $\mb z$ is a random input seed sampling from an uniform distribution, and $\ell$ is a loss function (e.g., mean squared error). After obtaining the optimal solution $\mb \theta^*$ of~\cref{eq:data_fitting_dip}, the clean image $\mb x$ could be easily restored with a forward pass $G_{\mb \theta^*} (\mb z)$. Despite the fact that DIP is learned without any training dataset, it has shown tremendous promise in a variety of tasks ranging from classical image restoration~\cite{DBLP_conf_cvpr_UlyanovVL18,li2023deep,li_random_2023}, to advanced computational
imaging problems~\cite{DBLP_conf_cvpr_GandelsmanSI19,DBLP_conf_cvpr_RenZWHZ20,DBLP_journals_corr_abs220809483}, and even beyond (e.g., time series~\cite{DBLP_conf_acssc_RavulaD22}).


\subsection{Problem Formulation}
We represent the clean RAW image as $\mb x^{1ch} \in \R^{H\times W}$, noise as $\mb n \in \R^{H\times W}$, and the noisy RAW image as $\wh{\mb x}^{1ch} = \mb x^{1ch} + \mb n$, where $H$ and $W$ are the height and width of the image. We further define a mask $\mb m \in \R^{H\times W \times 3}$, where each spatial location has $1$ in the channel corresponding to the color acquired at that position in the RGGB Bayer pattern, $0$ in other channels. We also introduce an operation $\mathcal{T}$, which maps the single-channel RAW data into a full-color image, where the non-Bayer components are $0$ (representing the missing pixels) and the originally sampled RAW data placed at the respective Bayer location and channels.

The goal of JDD is to reconstruct a high-quality full-color RGB image $\mb x^{3ch} \in \R^{H \times W \times 3}$ from a single noisy observation $\wh{\mb x}^{3ch} = \mathcal{T}\wh{\mb x}^{1ch}$. To do so, we need to fill in the missing pixels in the RAW image $\wh{\mb x}^{3ch}$ (\emph{demosaicing}), as well as to remove noisy components $\mb n$ (\emph{denoising}). In a manner similar to V-DIP~\cite{DBLP_journals_sensors_ParkLJY20}, we also formulate the demosaicing procedure as an image-inpainting problem, since both attempt to reconstruct the absent pixels.

\subsection{Double DIPs for Joint Demosaicing and Denoising}

DIP can restore high-quality images for inpainting when the observation is clean~\cite{DBLP_conf_cvpr_UlyanovVL18,li2023deep}, however, its performance deteriorates dramatically if additional noise is present~\cite{DBLP_conf_bmvc_LiZLPWS21}. Thus, we posit that conceptualizing the demosaicing procedure as an inpainting task and employing a DIP to directly restore absent values, without meticulous consideration of noise removal, may lead to suboptimal solutions.

To overcome this limitation, we introduce a novel dual-branch model for demosaicing (DM) and denoising (DN), dubbed~\emph{JDD-DoubleDIP}, which consists of two DIPs (see~\cref{fig_framework}):~\emph{DIP-Denoising} $G_{\mb \theta}$, parameterized by $\mb \theta$ and~\emph{DIP-Demosaicing}\footnote{We refer to this branch as DIP-Demosaicing simply to differentiate it from the denoising branch, but we note that it actually performs \emph{joint} demosaicing and denoising as its reconstruction is based on both the given noisy RAW image and the denoised output of DIP-Denoising branch.} $G_{\mb \phi}$, parameterized by $\mb \phi$. Specifically,~\emph{DIP-Denoising} $G_{\mb \theta}$ is dedicated to remove the noisy components in the given noisy single-channel RAW image $\wh{\mb x}^{1ch}$. Its loss function thus can be formulated as~\cref{eq_DN_loss}:
\begin{align} \label{eq_DN_loss}
   \ell_{\text{DN}} = \ell( \wh{\mb x}^{1ch}, G_{\mb \theta}(\mb z_{\text{DN}})) 
\end{align}
where $\mb z_{\text{DN}}$ is a random seed sampled from a normal distribution and $\ell$ is a distance measurement function (e.g., mean squared error).

\emph{DIP-Demosaicing} $G_{\mb \phi}$, on the other hand, is tailored to generate the desired high-quality full-color RGB image $\mb x^{3ch}$. Unlike V-DIP~\cite{DBLP_journals_sensors_ParkLJY20}, which reconstructs ${\mb x}^{3ch}$ only based on the given noisy RAW image $\wh{\mb x}^{3ch}$, our~\emph{DIP-Demosaicing} accomplishes this goal by making use of two sources of information---the given noisy RAW image $\wh{\mb x}^{3ch}$ and the reconstructed result of the~\emph{DIP-Denoising}. Therefore, the loss function of~\emph{DIP-Demosaicing} consists of two components, as shown in~\cref{eq_DM_loss}:
\begin{equation}
\begin{aligned} \label{eq_DM_loss}
   \ell_{\text{DM}} = &\ell(\mathcal{T} G_{\mb \theta}(\mb z_{\text{DN}})\odot \mb m, G_{\mb \phi}(\mb z_{\text{DM}})\odot \mb m)\\
   &+ \alpha\ell( \wh{\mb x}^{3ch}\odot \mb m, G_{\mb \phi}(\mb z_{\text{DM}})\odot \mb m)  
\end{aligned}
\end{equation}
where the first term simulates the procedure of using~\emph{DIP-Demosaicing} to demosaic a denoised (``clean") RAW image while the second term accounts for the original noisy RAW image demosaicing. We use $\alpha$ to weight the second loss term. Similarly, the input $\mb z_{\text{DM}}$ is sampled from a normal distribution and $\ell$ is the mean squared error.

Finally, we couple~\emph{DIP-Denoising} and~\emph{DIP-Demosaicing} toghter with a joint loss function, as shown in~\cref{eq_Joint_loss}:
\begin{align} \label{eq_Joint_loss}
   \ell_{\text{JDD-DoubleDIP}} = \sqrt{\ell_{\text{DN}}} + \sqrt{\ell_{\text{DM}}}
\end{align}
We use the square root of $\ell_{\text{DN}}$ and $\ell_{\text{DM}}$ to further suppress the effects of noise. We hypothesize that~\emph{DIP-Denoising} offers a form of ``clean" guidance to~\emph{DIP-Demosaicing}, enabling the latter to demosaic a ``clean" RAW image; conversely,~\emph{DIP-Demosaicing} serves as a mechanism for providing ``feedback" to~\emph{DIP-Denoising}, facilitating the removal of noisy components. Therefore, by training these two neural networks jointly, we expect to see improved performance of demosaicing and denoising.

\begin{figure*}[t]
\centering
\includegraphics[width=1\textwidth]{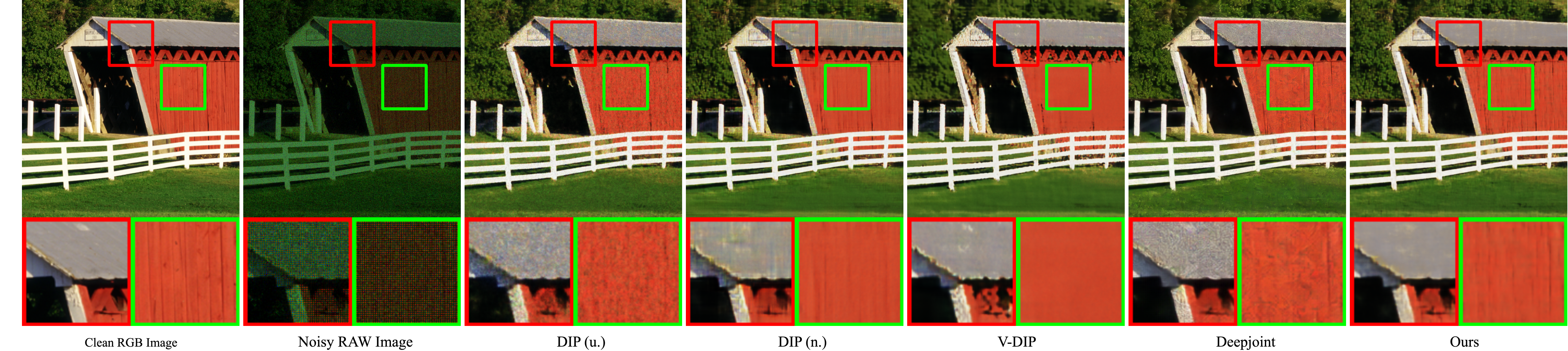}
\vspace{-20pt}
\caption{The visual comparisons of demosaicing and denoising on the RAW image (McMaster18) with Poisson noise ($\lambda=45$).}\vspace{-5pt}
\label{fig_vis}
\end{figure*}

\begin{table*}[!htbp]
\caption{The quantitative results of demosaicing and denoising of RAW images with Poisson noise. \textbf{Deepjoint}: we use the maximum noise intensity of the pretrained model as input. \textbf{$\text{Ours}^{+}$}: we smooth the output of our model by using running average: $\mb x^{3ch}_{\text{smooth}} = 0.99*\mb x^{3ch}_{\text{smooth}} + 0.01*G_{\phi}(\mb z_{\text{DM}}).$} 
\vspace{-15pt}
\label{tab_poisson_noise}
\begin{center}
\setlength{\tabcolsep}{1.15mm}{

\begin{tabular}{c l l l l l l| l l l l l}

&
& 
\multicolumn{5}{c}{\scriptsize{PSNR $\uparrow$}} &
\multicolumn{5}{c}{\scriptsize{SSIM $\uparrow$}}
\\
\hline

\hline

\hline
\scriptsize{Dataset}

& \scriptsize{Method}

& \scriptsize{$\lambda$=65}
& \scriptsize{$\lambda$=45}
& \scriptsize{$\lambda$=25}
& \scriptsize{$\lambda$=15}
& \scriptsize{$\lambda$=5}

& \scriptsize{$\lambda$=65}
& \scriptsize{$\lambda$=45}
& \scriptsize{$\lambda$=25}
& \scriptsize{$\lambda$=15}
& \scriptsize{$\lambda$=5}
\\
\hline

\hline
\multirow{7}{*}{\rotatebox{90}{\scriptsize{Kodak}}}

&\scriptsize{DIP (u.)}
&\scriptsize{26.184} \tiny{(0.018)}
&\scriptsize{25.429} \tiny{(0.024)}
&\scriptsize{24.181} \tiny{(0.023)}
&\scriptsize{23.175} \tiny{(0.025)}
&\scriptsize{21.069} \tiny{(0.031)}

&\scriptsize{0.675} \tiny{(0.0013)}
&\scriptsize{0.637} \tiny{(0.0015)}
&\scriptsize{0.5766} \tiny{(0.0011)}
&\scriptsize{0.5301} \tiny{(0.0018)}
&\scriptsize{0.4506} \tiny{(0.0028)}
\\

&\scriptsize{DIP (n.)}
&\scriptsize{27.349} \tiny{(0.029)}
&\scriptsize{26.662} \tiny{(0.038)}
&\scriptsize{25.537} \tiny{(0.016)}
&\scriptsize{24.584} \tiny{(0.021)}
&\scriptsize{22.297} \tiny{(0.016)}

&\scriptsize{0.745} \tiny{(0.0005)}
&\scriptsize{0.716} \tiny{(0.0010)}
&\scriptsize{0.668} \tiny{(0.0005)}
&\scriptsize{0.625} \tiny{(0.0014)}
&\scriptsize{0.519} \tiny{(0.0008)}
\\

&\scriptsize{V-DIP}
&\scriptsize{27.131} \tiny{(0.043)}
&\scriptsize{26.382} \tiny{(0.053)}
&\scriptsize{25.253} \tiny{(0.026)}
&\scriptsize{24.174} \tiny{(0.039)}
&\scriptsize{21.910} \tiny{(0.023)}

&\scriptsize{0.756} \tiny{(0.0011)}
&\scriptsize{0.730} \tiny{(0.0018)}
&\scriptsize{0.687} \tiny{(0.0002)}
&\scriptsize{0.645} \tiny{(0.0021)}
&\scriptsize{0.547} \tiny{(0.0017)}
\\





&\scriptsize{Deepjoint}
&\scriptsize{28.883} \tiny{(N/A)}
&\scriptsize{26.850} \tiny{(N/A)}
&\scriptsize{22.504} \tiny{(N/A)}
&\scriptsize{19.366} \tiny{(N/A)}
&\scriptsize{14.728} \tiny{(N/A)}

&\scriptsize{0.770} \tiny{(N/A)}
&\scriptsize{0.682} \tiny{(N/A)}
&\scriptsize{0.487} \tiny{(N/A)}
&\scriptsize{0.347} \tiny{(N/A)}
&\scriptsize{0.168} \tiny{(N/A)}
\\

&\scriptsize{Ours}
&\scriptsize{28.116} \tiny{(0.029)}
&\scriptsize{27.436} \tiny{(0.036)}
&\scriptsize{26.257} \tiny{(0.041)}
&\scriptsize{25.248} \tiny{(0.029)}
&\scriptsize{22.750} \tiny{(0.018)}

&\scriptsize{0.780} \tiny{(0.0015)}
&\scriptsize{0.754} \tiny{(0.0008)}
&\scriptsize{0.711} \tiny{(0.0009)}
&\scriptsize{0.676} \tiny{(0.0009)}
&\scriptsize{0.593} \tiny{(0.0018)}
\\

&\scriptsize{$\text{Ours}^{+}$}
&\scriptsize{28.515} \tiny{(0.040)}
&\scriptsize{27.820} \tiny{(0.035)}
&\scriptsize{26.593} \tiny{(0.036)}
&\scriptsize{25.536} \tiny{(0.028)}
&\scriptsize{22.844} \tiny{(0.020)}

&\scriptsize{0.794} \tiny{(0.0009)}
&\scriptsize{0.769} \tiny{(0.0005)}
&\scriptsize{0.726} \tiny{(0.0007)}
&\scriptsize{0.690} \tiny{(0.0013)}
&\scriptsize{0.602} \tiny{(0.0012)}
\\
\hline

\hline
\multirow{7}{*}{\rotatebox{90}{\scriptsize{McMaster }}}

&\scriptsize{DIP (u.)}
&\scriptsize{26.610} \tiny{(0.015)}
&\scriptsize{25.774} \tiny{(0.040)}
&\scriptsize{24.431} \tiny{(0.041)}
&\scriptsize{23.412} \tiny{(0.040)}
&\scriptsize{21.242} \tiny{(0.028)}

&\scriptsize{0.722} \tiny{(0.0012)}
&\scriptsize{0.686} \tiny{(0.0031)}
&\scriptsize{0.627} \tiny{(0.0028)}
&\scriptsize{0.583} \tiny{(0.0023)}
&\scriptsize{0.501} \tiny{(0.0046)}
\\

&\scriptsize{DIP (n.)}
&\scriptsize{28.074} \tiny{(0.031)}
&\scriptsize{27.311} \tiny{(0.024)}
&\scriptsize{26.064} \tiny{(0.018)}
&\scriptsize{24.940} \tiny{(0.008)}
&\scriptsize{22.329} \tiny{(0.011)}

&\scriptsize{0.791} \tiny{(0.0007)}
&\scriptsize{0.766} \tiny{(0.0009)}
&\scriptsize{0.720} \tiny{(0.0012)}
&\scriptsize{0.676} \tiny{(0.0009)}
&\scriptsize{0.570} \tiny{(0.0009)}
\\

&\scriptsize{V-DIP}
&\scriptsize{27.908} \tiny{(0.040)}
&\scriptsize{26.990} \tiny{(0.028)}
&\scriptsize{25.664} \tiny{(0.053)}
&\scriptsize{24.535} \tiny{(0.009)}
&\scriptsize{22.004} \tiny{(0.015)}

&\scriptsize{0.801} \tiny{(0.0012)}
&\scriptsize{0.776} \tiny{(0.0002)}
&\scriptsize{0.733} \tiny{(0.0017)}
&\scriptsize{0.692} \tiny{(0.0009)}
&\scriptsize{0.594} \tiny{(0.0019)}
\\





&\scriptsize{Deepjoint}
&\scriptsize{29.149} \tiny{(N/A)}
&\scriptsize{27.210} \tiny{(N/A)}
&\scriptsize{23.472} \tiny{(N/A)}
&\scriptsize{20.590} \tiny{(N/A)}
&\scriptsize{15.748} \tiny{(N/A)}

&\scriptsize{0.788} \tiny{(N/A)}
&\scriptsize{0.715} \tiny{(N/A)}
&\scriptsize{0.595} \tiny{(N/A)}
&\scriptsize{0.497} \tiny{(N/A)}
&\scriptsize{0.299} \tiny{(N/A)}
\\

&\scriptsize{Ours}
&\scriptsize{28.805} \tiny{(0.017)}
&\scriptsize{28.056} \tiny{(0.022)}
&\scriptsize{26.836} \tiny{(0.023)}
&\scriptsize{25.698} \tiny{(0.045)}
&\scriptsize{23.029} \tiny{(0.031)}

&\scriptsize{0.813} \tiny{(0.0014)}
&\scriptsize{0.792} \tiny{(0.0010)}
&\scriptsize{0.753} \tiny{(0.0012)}
&\scriptsize{0.717} \tiny{(0.0019)}
&\scriptsize{0.632} \tiny{(0.0037)}
\\

&\scriptsize{$\text{Ours}^{+}$}
&\scriptsize{29.286} \tiny{(0.015)}
&\scriptsize{28.510} \tiny{(0.019)}
&\scriptsize{27.222} \tiny{(0.017)}
&\scriptsize{26.019} \tiny{(0.040)}
&\scriptsize{23.123} \tiny{(0.036)}

&\scriptsize{0.825} \tiny{(0.0014)}
&\scriptsize{0.804} \tiny{(0.0010)}
&\scriptsize{0.766} \tiny{(0.0013)}
&\scriptsize{0.730} \tiny{(0.0021)}
&\scriptsize{0.641} \tiny{(0.0030)}
\\
\hline

\hline

\hline
\end{tabular}
}
\end{center}\vspace{-24pt}
\end{table*}

\section{Experiments}\label{sec:expriments}


\subsection{Experimental Settings}
We modify the original DIP~\cite{DBLP_conf_cvpr_UlyanovVL18} architecture slightly for implementing~\emph{DIP-Denoising} and~\emph{DIP-Demosaicing}: 1) the input seed $\mb z_{\text{DN}}$ and $\mb z_{\text{DM}}$ are sampled from a normal distribution $\mathcal{N}(0, 0.01)$; 2) the output of~\emph{DIP-Denoising} is a single-channel image while the output of~\emph{DIP-Demosaicing} is a three-channel image. 
To verify the effectiveness of our method, we conduct experiments on two popular image datasets---Kodak~\cite{Kodak_dataset} and McMaster~\cite{DBLP_journals_jei_0006WB011}. We also experiment with two common sensor noises---Gaussian noise and Poisson noise---on different noise intensities, ranging from benign
to severe scenarios. 
We compare our method with three single-instance methods---DIP (u.), DIP (n.)~\cite{DBLP_conf_cvpr_UlyanovVL18}, V-DIP~\cite{DBLP_journals_sensors_ParkLJY20}; as well as one data-driven method---Deepjoint~\cite{DBLP_journals_tog_Durand16a}. DIP (u.) indicates that the input $\mb z$ is sampled from a uniform distribution $\mathcal{U}(0, 0.1)$, while in DIP (n.) the input $\mb z$ is sampled from a normal distribution $\mathcal{N}(0, 0.01)$. 
For the implementation of V-DIP, we adhere to its original paper~\cite{DBLP_journals_sensors_ParkLJY20}. 
For Deepjoint, we employ its official pretrained model which was trained on Gaussian noise intensities $\sigma \in [0, 20]$. For our own model, we fix hyperparameters for all experiments: we set $\alpha$ to $0.1$, the learning rate of~\emph{DIP-Denoising} to $5\times10^{-3}$, the learning rate of~\emph{DIP-Demosaicing} to $5\times 10^{-2}$, and we use Adam~\cite{DBLP_journals_corr_KingmaB14} as our optimizer. We adopt PSNR and SSIM as our performance metrics. To ensure the consistency of our observations, we repeat experiments for $5$ rounds and then report their mean and standard deviation.


\subsection{RAW Images with Gaussian Noise}\label{subsec_Gaussian_noise}

Here, we first experiment with RAW images corrupted by Gaussian noise. We simulate a noisy RAW image $\wh{\mb x} = \mb x + \mb n$ where $\mb n = \mathcal{N} (0, \sigma)$. 
$\sigma$ is varied within $\{10, 20, 30, 50, 70\}$ to account for practical scenarios with different severity.

~\cref{tab_gaussin_noise} shows the quantitative comparisons. One can observe that: 1) our method consistently outperforms other single- instance methods---DIP (u.), DIP (n.), and V-DIP---on both datasets and across all noise intensities; 2) compared with the
data-driven method---Deepjoint\footnote{\textbf{Deepjoint}: input the maximum noise intensity of the pretrained model.}
and $\text{Deepjoint}^{*}$\footnote{\textbf{$\text{Deepjoint}^{*}$}: input the exact noise intensity of the noisy RAW image.}, when the noise intensity is in the distribution of the pretrained model
(e.g., $\sigma\in\{10, 20\}$) , Deepjoint performs slighlty better than our method and $\text{Deepjoint}^{*}$ further improves the 
performance of Deepjoint; while when the test noise intensity is 
out of the distribution of the pretrained model (e.g., $\sigma \in \{30, 50, 70\}$), the superiority of our approach becomes apparent, 
even compared with the $\text{Deepjoint}^{*}$ setting, 
thereby validating the distribution shift issue of data-driven methods; 
3) applying a running average smoothing on the outputs of our method, which is indicated as $\text{Ours}^{+}$, further improves the performance.

\begin{table}[!htpb]
\caption{The comparison of demosaicing and denoising of JDD-DoubleDIP and Over-Parameterization.}
\vspace{-10pt}
\label{tab_overparam}
\begin{center}
\setlength{\tabcolsep}{2.25mm}{

\begin{tabular}{c l l | l l}

&
\multicolumn{2}{c}{\scriptsize{Gaussian Noise}} &
\multicolumn{2}{c}{\scriptsize{Poisson Noise}}
\\
\hline

\hline

\hline

\scriptsize{Method}

& \scriptsize{\scriptsize{PSNR $\uparrow$}}
& \scriptsize{\scriptsize{SSIM $\uparrow$}}
& \scriptsize{\scriptsize{PSNR $\uparrow$}}
& \scriptsize{\scriptsize{SSIM $\uparrow$}}
\\
\hline


\scriptsize{DM-DM}
&\scriptsize{26.226} \tiny{(0.246)}
&\scriptsize{0.711} \tiny{(0.0146)}
&\scriptsize{26.128} \tiny{(0.199)}
&\scriptsize{0.737} \tiny{(0.0112)}
\\

\scriptsize{Ours}
&\scriptsize{26.880} \tiny{(0.013)}
&\scriptsize{0.726} \tiny{(0.0013)}
&\scriptsize{26.836} \tiny{(0.023)}
&\scriptsize{0.753} \tiny{(0.0012)}
\\
\hline

\hline

\hline
\end{tabular}
}
\end{center}\vspace{-25pt}
\end{table}

\subsection{RAW Images with Poisson Noise}\label{subsec_poisson_noise}

Now, we test our method on RAW images with Poisson noise, which is another dominant noise in camera sensors, especially in scenarios with low-light conditions. We simulate the pixel-wise independent Poisson noise as follows: for each pixel $x \in [0,1]$, the noisy pixel is Poisson distributed with rate $\lambda x$ and we test different intensities of noise by varying $\lambda \in \{65, 45, 25, 15, 5\}$. We report the experimental results in~\cref{tab_poisson_noise}, which is consistent with our observations in~\cref{subsec_Gaussian_noise}, reassuring the effectiveness of our method. 

We also depict a visual comparison in~\cref{fig_vis}. It is evident that the reconstructions by other methods introduce some artifacts (e.g., residual noise and distortion artifacts), while our reconstruction preserves more texture details and is more perceptually pleasing, reinforcing the benefits of our method.

\subsection{JDD-DoubleDIP vs. Over-Parameterization}

To verify that the benefits of our method are not a consequence of over-parameterization owing to the use of two DIPs in conjunction, we design a counterpart for our method named \emph{DM-DM} in which we replace~\emph{DIP-Denoising} with another~\emph{DIP-Demosaicing}, resulting in a network parameterized similarly to ours. For simplicity, here we only experiment with McMaster on Gaussian noise ($\sigma=30$) and Poisson noise ($\lambda=25$). The experimental results are reported in~\cref{tab_overparam}. It is evident that our method yields higher PSNR and SSIM on both noises, and our method is much more stable ( our method has $>10\times$ lower standard deviation compared with~\emph{DM-DM}), suggesting that the benefits of our method are likely not due to over-parameterization.

\section{Conclusion}\label{sec:conclusion}

In this paper, we propose a novel joint demosaicing and denoising method, dubbed~\emph{JDD-DoubleDIP}, which consists of a denoising DIP to explicitly account for noise removal and a demosaicing DIP for high-quality full-color RGB image generation. By training these two DIPs jointly, our method yields better PSNR, SSIM, and perceived visual quality on Kodak and McMaster datasets under different noise types and intensities compared to other methods. 
In the future, we would like to investigate the explicit incorporation of prior information from the green channel into our method to boost performance, as it has twice as much information as the red or blue channels in a RAW image.




\bibliographystyle{IEEEbib}
\bibliography{doubleDIP}

\end{document}